\documentclass[preprint,10pt,sort&compress]{elsarticle}
\usepackage{amssymb,amsfonts,amsmath}
\usepackage{graphicx}
\pdfoutput=1

\begin{document}
\renewcommand{\thefootnote}{\fnsymbol{footnote}}
\begin{center} 
{\Large \bf Dependence of nuclear spin singlet lifetimes on RF spin-locking power}
\end{center}

\noindent Stephen J. DeVience$^{a}$*, Ronald L. Walsworth$^{b,c}$, Matthew S. Rosen$^{c,d,e}$\\ \\
$^{a}$ Department of Chemistry and Chemical Biology, Harvard University, 12 Oxford St., Cambridge, MA 02138\\
$^{b}$ Harvard-Smithsonian Center for Astrophysics, 60 Garden St., Cambridge, MA 02138\\
$^{c}$ Department of Physics, Harvard University, 17 Oxford St., Cambridge, MA 02138\\
$^{d}$ Harvard Medical School, 25 Shattuck Street, Boston, MA 02115\\
$^{e}$ Martinos Center for Biomedical Imaging, 149 Thirteenth St., Charlestown, MA 02129\\

*Corresponding Address:\\
Harvard-Smithsonian Center for Astrophysics,\\
MS 59, 60 Garden St., Cambridge, MA 02138\\
email: devience@fas.harvard.edu

\section*{Abstract}
We measure the lifetime of long-lived nuclear spin singlet states as a function of the strength of the RF spin-locking field and present a simple theoretical model that agrees well with our measurements, including the low-RF-power regime. We also measure the lifetime of a long-lived coherence between singlet and triplet states that does not require a spin-locking field for preservation. Our results indicate that for many molecules, singlet states can be created using weak RF spin-locking fields: more than two orders of magnitude lower RF power than in previous studies.  Our findings suggest that in many biomolecules, singlets and related states with enhanced lifetimes might be achievable {\em in vivo} with safe levels of RF power.

\section{Introduction}
The spin-lattice relaxation time $T_1$ is a limiting factor for a broad class of NMR experiments in which nuclear spin polarization or order needs to be preserved or transported \cite{Golman1, Golman2, Yu1, Callaghan1}. However, long-lived nuclear spin singlet states with lifetimes up to 37 $T_1$ have recently been measured in thermally-polarized samples \cite{Levitt1,Levitt2,Levitt6,Bodenhausen4,Bodenhausen5,Warren1}. These long-lived states have been used to study slow processes such as diffusion, chemical exchange, and conformational dynamics {\em in vitro} \cite{Bodenhausen3, Bodenhausen6, Bodenhausen2}. 

Nuclear spin singlet states are typically created from pairs of coupled spin-1/2 nuclei with equal or near-equal resonance frequencies. The most well-known example is the H$_2$ molecule, which exists in two forms: para-H$_2$, with the singlet eigenstate $ \vert S_0 \rangle = (\vert\uparrow \downarrow \rangle - \vert\downarrow \uparrow \rangle)/\sqrt{2}$; and ortho-H$_2$, with triplet eigenstates $ \vert T_{-} \rangle = \vert\uparrow \uparrow \rangle$, $ \vert T_0 \rangle = (\vert\uparrow \downarrow \rangle + \vert\downarrow \uparrow \rangle)/\sqrt{2}$, and $ \vert T_{+} \rangle = \vert\downarrow \downarrow \rangle$ \cite{Farkas1}. In our notation $\uparrow$ represents a spin aligned with the applied magnetic field, $B_0$, while $\downarrow $ represents a spin that is anti-aligned. In the singlet state, the net nuclear spin is zero, and there is no net magnetic dipole moment. Hence interactions with the environment are weak and the rate of interconversion between singlet and triplet states is very slow, often much slower than the spin-lattice relaxation rate $1/T_1$. On the other hand, the triplet states have non-zero magnetic moments and couple strongly with the environment. Relaxation among the triplet states occurs on the timescale $T_1$.

Recently, Levitt and colleagues demonstrated a general technique for the creation of singlet states for pairs of magnetically inequivalent nuclei \cite{Levitt6,Levitt9}. Such inequivalent nuclear spins cannot form ideal, long-lived singlet states as outlined above: the different local environments of the nuclear spins leads to rapid conversion to the triplet state and thus coupling to the environment. Nonetheless, as Levitt {\em et al.} showed, a properly designed RF pulse sequence can prepare a singlet state, which is then preserved from triplet interconversion by the application of a continuous resonant RF field. This ``spin-locking'' field forces the average Hamiltonian of the two nuclear spins to be effectively equivalent. While this technique is applicable for a large variety of molecules, the large continuous RF power employed in spin-locking experiments to date implies an RF specific absorption rate (SAR) that is likely prohibitive for animal and human studies \cite{Bottomley1,IEC1}.

In this paper, we report measurements of singlet state lifetime for a variety of organic molecules and as a function of RF spin-locking field strength. We find that the measured RF power required to preserve a singlet state agrees well with the predictions of a simple theoretical model with inputs from the molecule's NMR spectrum. We also present measurements of a singlet-triplet coherence with an extended lifetime that does not require the use of RF spin locking for preservation. Moreover, our findings demonstrate that for many molecules of interest, singlet lifetimes many times longer than $T_1$ can be achieved with much weaker RF spin-locking fields than have been used to date -- more than an order of magnitude smaller than in previous studies -- leading to both an RF power and an SAR more than 100 times lower. This result suggests that {\em in vivo} application of long-lived singlet NMR might be possible in biomolecules with the appropriate properties, despite limitations imposed by RF SAR.

\section{Theory}
\subsection{Singlet Relaxation Mechanisms}

Many relaxation mechanisms are forbidden by symmetry from converting the singlet state to triplet states. For example, magnetic dipole-dipole interactions between the singlet's two spins cannot couple the antisymmetric singlet state to the symmetric triplet states. Since this intra-pair interaction is often the dominant driver of relaxation, the typical result is a singlet population with a lifetime $T_S$ many times longer than the spin-lattice relaxation time $T_1$. Dipole-dipole interactions between the singlet spin pair and more distant spins can also lead to relaxation, but the singlet is protected from dipolar fluctuations common to both singlet spins: i.e., in the far-field the net dipole moment of the singlet is zero. Thus singlet-state relaxation must instead occur through differential interactions on each spin of the singlet; these include chemical shift anisotropy (CSA), spin rotation (SR) due to collisions between molecules, and magnetic quadrupole interactions with a third spin \cite{Pileio1,Grant1,Grant2,Levitt10}. 

Since these three singlet relaxation mechanisms respond differently to the applied static magnetic field and temperature, the dominant effect can be determined experimentally. CSA has a strong magnetic-field dependence with a lifetime scaling as $T_{CSA} \propto 1/B_0^2$ \cite{Pileio1}. Spin rotation collisions result in a lifetime scaling non-linearly with temperature as $T_{SR} \propto exp(E/k_B T) \propto \frac{E}{k_B T}$ \cite{Pileio1}. Magnetic quadrupole interactions, on the other hand, produce a singlet lifetime scaling linearly with temperature in the extreme-narrowing regime (when molecular rotation rates are much greater than the Larmor frequency). As demonstrated below, for the molecules used in the present study, the quadrupolar mechanism dominates singlet-state relaxation, i.e., $T_S \approx T_Q$.

Magnetic quadrupole relaxation results from the two spins of the singlet interacting differently with a third spin. This relaxation mechanism was modeled at high magnetic field by Tayler \emph{et al.} \cite{Levitt10}, who derived an expression for the enhancement of the singlet lifetime\footnote{Ref. 21 contains a typesetting error in which the summation has been taken over the whole expression rather than only the denominator.}:
\begin{equation}
\frac{T_S}{T_1} = \frac{3b_{12}^2}{2\displaystyle\sum\limits_{j>2}(b^2_{1j}+b^2_{2j}-b_{1j} b_{2j} (3\cos^2 \phi_{1j2}-1))}   .
\end{equation}
Here spins 1 and 2 compose the singlet while $j$ represents another nearby spin; $b_{jk} = \gamma^2 / r_{jk}^3 $ is a measure of the dipolar coupling strength between spins; and $\phi_{1j2} $ is the angle between the vectors connecting 1 with $j$ and 2 with $j$. In principle, there is no limit to the singlet lifetime enhancement given the proper molecular geometry. However, in practice other relaxation mechanisms gain importance if magnetic quadrupole relaxation is highly suppressed. In a previous study, Equation 1 was found to agree well with measurements of singlet-state lifetimes using high RF spin-locking power \cite{Levitt10}. 

\subsection{RF-Power Dependence}

\begin{figure*}
\vspace*{.05in}
\centering
\includegraphics[scale=1]{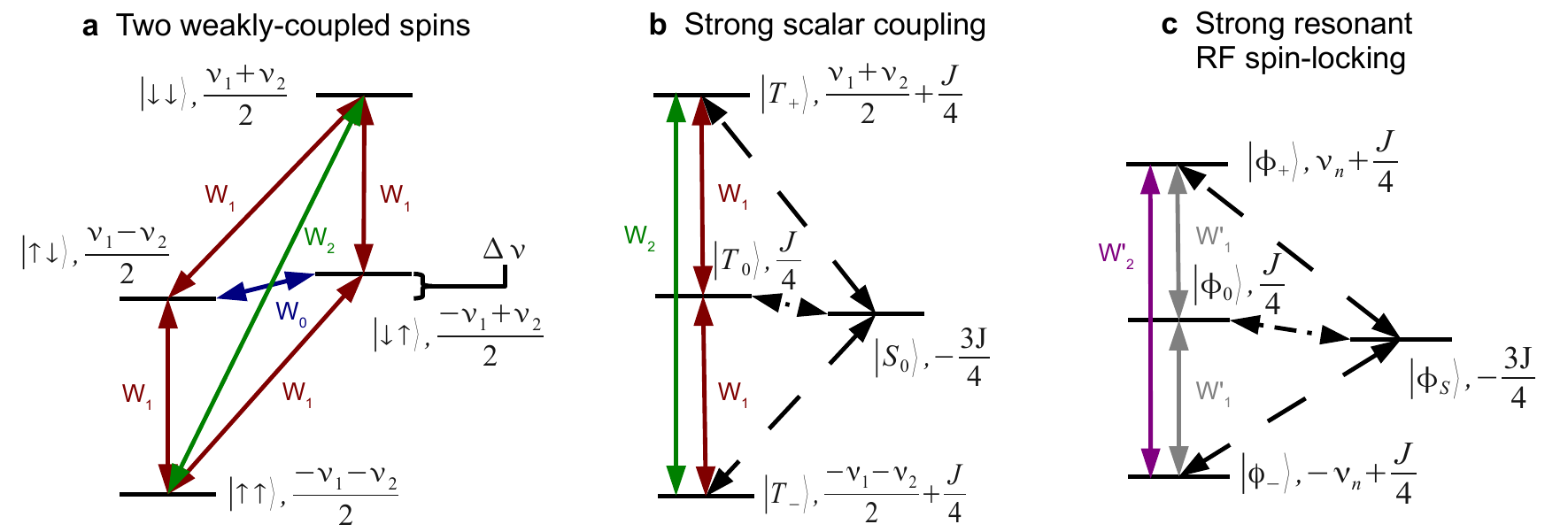}  
\caption{A system containing two spins with resonant transition frequencies $\nu_1$ and $\nu_2$ can be represented by four spin-pair eigenstates. (A) For weak spin coupling, relaxation occurs via magnetic dipole-dipole interactions due to zero-, single-, and double-quantum transitions (with rates $W_0$, $W_1$, and $W_2$). (B) If the two spins are coupled strongly by scalar coupling (J $>> \Delta \nu$, the chemical shift splitting), the spin-pair eigenstates are mixed into singlet and triplet states. The triplet states still interact via dipole-dipole interactions (solid lines), but singlet-triplet transitions are forbidden (dashed lines). (C) Mixing of the spin-pair eigenstates can also be induced by a strong RF spin-locking field ($\nu_n > 5\Delta \nu$), which produces a singlet state and three mixtures of triplet states. Singlet-triplet transitions are again forbidden. The new triplet states exhibit new transition rates $W'_1$ and $W'_2$.}

\end{figure*}

A detailed theoretical analysis of the singlet's lifetime during RF irradiation has been given by Pileio and Levitt, who performed exact numerical calculations for the relationship between singlet lifetime and RF power \cite{Pileio1}. Here, we develop an approximate model that leads to a simple calculation of the measured singlet lifetime at a given spin-locking frequency. Our model can easily be fit to measurements of singlet lifetime at a number of RF field strengths so that the maximum singlet lifetime can be extracted.

A pair of spin-1/2 nuclei creates a system with four spin-pair eigenstates as shown in Fig. 1A: $\vert \uparrow \uparrow \rangle, \vert \uparrow \downarrow \rangle, \vert \downarrow \uparrow \rangle$, and $\vert \downarrow \downarrow \rangle$. In general, there is mixing between the two eigenstates with no net z-component of spin ($\vert \uparrow \downarrow \rangle$ and $\vert \downarrow \uparrow \rangle$) if the spin states are degenerate, or if scalar coupling between the spins is strong compared with any difference between the individual spin transition frequencies ($\nu_1$ and $\nu_2$). As a result of such mixing, the spin-pair system is described by one singlet state, $ \vert S_0 \rangle = (\vert \uparrow \downarrow \rangle - \vert \downarrow \uparrow \rangle)/\sqrt{2}$, and three triplet states $ \vert T_{-} \rangle = \vert \uparrow \uparrow \rangle$, $ \vert T_0 \rangle = (\vert \uparrow \downarrow \rangle + \vert \downarrow \uparrow \rangle)/\sqrt{2}$, and $ \vert T_{+} \rangle = \vert \downarrow \downarrow \rangle$, with the energy levels shown in Fig. 1B. In this case, spin polarization cannot be transfered to the singlet state from the triplet states via an RF pulse sequence because $\mathcal{H}_{RF} \vert S_0 \rangle = 0 $.

For many molecules of interest, however, chemical shifts induce a difference between spin transition frequencies ($\Delta \nu = \vert \nu_1-\nu_2 \vert$) that is much larger than the scalar coupling, and hence there is little mixing of the bare spin-pair eigenstates. In this case unitary transformations, via the RF pulse sequence shown in Fig. 2A, can transfer initial thermal spin polarization to the singlet state with at most 50\% efficiency by creating the singlet-enhanced superposition state \cite{Levitt9}
\begin{align}
\rho_{ST} =& \vert T_0 \rangle \langle T_0 \vert - \vert S_0 \rangle \langle S_0 \vert\\
 =& \vert \uparrow \downarrow \rangle \langle \downarrow \uparrow \vert + \vert \downarrow \uparrow \rangle \langle \uparrow \downarrow \vert.
\end{align}
Similarly, the RF pulse sequence shown in Fig. 2B can transfer initial thermal spin polarization into a long-lived coherence between the singlet and triplet states with density matrix
\begin{align}
\rho_{LLC} =& \vert S_0 \rangle \langle T_0 \vert + \vert T_0 \rangle \langle S_0 \vert\\
 =& \vert \uparrow \downarrow \rangle \langle \uparrow \downarrow \vert - \vert \downarrow \uparrow \rangle \langle \downarrow \uparrow \vert.
\end{align}
A similar long-lived coherence has been previously studied as a way to extend $T_2$ \cite{Bodenhausen1,Bodenhausen8}.

\begin{figure}
\vspace*{.05in}
\centering
\includegraphics[scale=1]{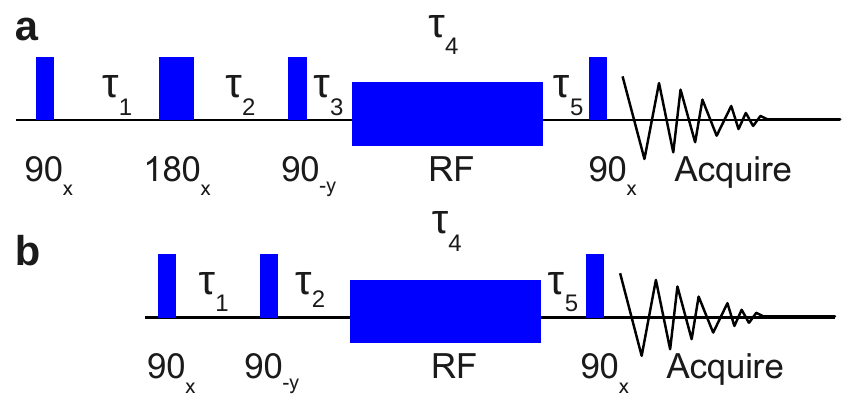}  
\caption{(A) The singlet-enhanced superposition state $\rho_{ST}$ is initialized via a 3-pulse preparation sequence. (B) A long-lived coherence (LLC) between the singlet and triplet states $\rho_{LLC}$, as well as mixtures of the singlet and LLC states, are initialized via a 2-pulse preparation sequence. Both preparation sequences are followed by application of the RF spin-locking field and a signal acquisition pulse. }
\end{figure}

For both $\rho_{ST}$ and $\rho_{LLC}$, population in the singlet state will be rapidly interconverted with the triplet states on a timescale $\sim 1/\Delta \nu$, providing strong coupling to the environment and thus rapid relaxation to the thermal state. However, a strong on-resonance RF field has been shown to be effective for maintaining the singlet-enhanced superposition state $\rho_{ST}$ \cite{Levitt6,Levitt9}. When the spin-locking RF field is set to the average resonant transition frequency of the two spins, the Hamiltonian in the bare spin-pair basis becomes
\begin{equation}
\mathcal{H} = h \left[ \begin{array}{cccc}
-\frac{\nu_1+\nu_2}{2} & \frac{\nu_n}{2} & \frac{\nu_n}{2} & 0 \\[0.5em]
\frac{\nu_n}{2} & \frac{\nu_2-\nu_1}{2} & 0 &\frac{\nu_n}{2}  \\[0.5em]
\frac{\nu_n}{2} & 0 & \frac{\nu_1-\nu_2}{2} & \frac{\nu_n}{2}  \\[0.5em]
0 & \frac{\nu_n}{2}  & \frac{\nu_n}{2}  & \frac{\nu_1+\nu_2}{2} \end{array} \right]   .
\end{equation}
Here $ \nu_n$ is the effective spin nutation frequency due to the RF field, which drives single-quantum spin transitions. Also, we assume scalar coupling is weak and neglect its contributions to the Hamiltonian.

Diagonalizing this Hamiltonian yields four spin-locked eigenstates given by 
\begin{align}
\vert \phi _{+}\rangle =& \frac{1}{2} \sin \theta (\vert \uparrow \uparrow \rangle+\vert \downarrow \downarrow \rangle)+\frac{1}{2} \cos \theta (\vert \uparrow \downarrow \rangle - \vert \downarrow \uparrow \rangle)\nonumber\\ +& \frac{1}{2}(\vert \uparrow \downarrow \rangle + \vert \downarrow \uparrow \rangle) \\
\vert \phi _{0}\rangle =& \frac{1}{\sqrt{2}}(\vert \uparrow \uparrow \rangle - \vert \downarrow \downarrow \rangle) \\
\vert \phi _{S}\rangle =& -\frac{1}{\sqrt{2}} \cos \theta (\vert \uparrow \uparrow \rangle+\vert \downarrow \downarrow \rangle)\nonumber\\ +& \frac{1}{\sqrt{2}} \sin \theta (\vert \uparrow \downarrow \rangle - \vert \downarrow \uparrow \rangle) \\
\vert \phi _{-}\rangle =& -\frac{1}{2} \sin \theta (\vert \uparrow \uparrow \rangle+\vert \downarrow \downarrow \rangle)-\frac{1}{2} \cos \theta (\vert \uparrow \downarrow \rangle - \vert \downarrow \uparrow \rangle)\nonumber\\ +& \frac{1}{2}(\vert \uparrow \downarrow \rangle + \vert \downarrow \uparrow \rangle)   .
\end{align}
The mixing angle $\theta$ is controlled by the ratio of the spin nutation frequency to the chemical shift splitting:
\begin{equation}
\theta  = \arctan \frac{2 \nu_n }{ \Delta \nu}   
\end{equation}

At very large nutation rates ($\nu_n >> \Delta \nu$), i.e., high RF spin-locking power, the spin-locked eigenstates simplify to 
\begin{align}
\vert \phi _{+}\rangle =& \frac{1}{2}(\vert \uparrow \downarrow \rangle + \vert \downarrow \uparrow \rangle+\vert \uparrow \uparrow \rangle + \vert \downarrow \downarrow \rangle)\nonumber\\ =& \frac{1}{\sqrt{2}}\vert T_0 \rangle + \frac{1}{2}(\vert T_{-} \rangle+\vert T_{+} \rangle)\\
\vert \phi _{0}\rangle =& \frac{1}{\sqrt{2}}(\vert \uparrow \uparrow \rangle - \vert \downarrow \downarrow \rangle)=\frac{1}{\sqrt{2}}(\vert T_{-} \rangle - \vert T_{+} \rangle)\\
\vert \phi _{S}\rangle =& \frac{1}{\sqrt{2}}(\vert \uparrow \downarrow \rangle - \vert \downarrow \uparrow \rangle) = \vert S_0 \rangle\\
\vert \phi _{-}\rangle =& \frac{1}{2}(\vert \uparrow \downarrow \rangle + \vert \downarrow \uparrow \rangle-\vert \uparrow \uparrow \rangle-\vert \downarrow \downarrow \rangle)\nonumber\\ =& \frac{1}{\sqrt{2}}\vert T_0 \rangle - \frac{1}{2}(\vert T_{-} \rangle+\vert T_{+} \rangle)   .
\end{align}
Note that the spin-locked singlet state $\vert \phi_S \rangle$ corresponds to $\vert S_0 \rangle$ in this limit of large spin nutation (i.e., large RF spin-locking field), whereas the three spin-locked triplet states are each mixtures of eigenstates $\vert T_0 \rangle,\vert T_{+} \rangle$ , and $\vert T_{-} \rangle$. In this case, the initial state $\rho_{ST}$ is described well by Equation 2. The singlet, $\vert S_0 \rangle$, is well-protected by the RF spin-locking field, and after a short initial period during which the triplet states equilibrate, the remaining $\vert S_0 \rangle$ component relaxes exponentially with the characteristic time predicted by Equation 1. 

In the high-RF-power regime, the long-lived coherence $\rho_{LLC}$ is a sum of coherences containing $\vert \phi_{+} \rangle$, $\vert \phi_{-} \rangle$, and $\vert \phi_S \rangle$, which experience decoherence due to both dipole-dipole interactions and inhomogeneities in the RF spin-locking field.

If instead very small RF spin-locking power is applied ($\nu_n << \Delta \nu$), the singlet component of $\rho_{ST}$ rapidly interconverts with the central triplet state, $\vert T_0 \rangle$. When no RF power is applied, $\rho_{ST}$ is a zero-quantum coherence that precesses in the xy-plane, with a lifetime up to 3.25 $T_1$ if inter-pair dipole-dipole interactions are the sole relaxation mechanism \cite{Ernst1}. The addition of a small amount of RF power quickly decreases the lifetime of the $\rho_{ST}$ coherence because the RF field efficiently drives single-quantum transitions but creates very little long-lived singlet component.

In the low-RF-power regime, the long-lived coherence $\rho_{LLC}$ is well-described by Equation 5 as a population difference between the two central bare spin-pair eigenstates. The conventional two-spin dipole-dipole relaxation model of Solomon \cite{Solomon1, Slichter1} predicts that in most cases $T_{LLC}$ = $3 T_{1}$ (see supplement S1).

For intermediate RF spin-locking power ($\nu_n \approx \Delta \nu$), a more complex analysis is required. For an arbitrary RF power, the initial state $\rho_{ST}$ can be represented as
\begin{align}
\rho_{ST} =& \textstyle\vert \uparrow \downarrow \rangle \langle \downarrow \uparrow \vert + \vert \downarrow \uparrow \rangle \langle \uparrow \downarrow \vert \nonumber\\ =& \frac{\cos^2 \theta}{2} (\vert \phi _{+}\rangle \langle \phi _{-}\vert-\vert \phi _{+}\rangle \langle \phi_{+} \vert +\vert \phi _{-}\rangle \langle \phi _{+}\vert-\vert \phi _{-}\rangle \langle \phi _{-}\vert)\nonumber\\ +&\frac{\cos \theta \sin \theta}{\sqrt{2}} (\vert \phi _{-}\rangle \langle \phi _{S}\vert-\vert \phi _{+}\rangle \langle \phi _{S}\vert +\vert \phi _{S}\rangle \langle \phi _{-}\vert-\vert \phi _{S}\rangle \langle \phi _{+}\vert)\nonumber\\ +&\frac{1}{2}(\vert \phi _{+}\rangle \langle \phi _{+}\vert+\vert \phi _{+}\rangle \langle \phi _{-}\vert +\vert \phi _{-}\rangle \langle \phi _{+}\vert+\vert \phi _{-}\rangle \langle \phi _{-}\vert)\nonumber\\  -& \sin^2 \theta \vert \phi _{S}\rangle \langle \phi_S \vert   ,
\end{align}

where the four spin-locked eigenstates are given by Eq. (7)-(10).

At moderate RF powers ($\nu_n > \Delta \nu$), $\rho_{ST}$ is still mainly composed of the population $\vert \phi_S \rangle \langle \phi_S \vert$ and mixed triplet states. However, the eigenstate $\vert \phi_S \rangle$ no longer consists solely of the singlet $\vert S_0 \rangle$. It also contains a triplet component $\cos \theta (\vert T_{-} \rangle + \vert T_{+} \rangle)/\sqrt{2} $, which interacts with $\vert \phi_0 \rangle $ via a double-quantum transition, with relaxation rate scaling as $\cos^2 \theta $. The triplet component also interacts with $\vert \phi_{+} \rangle $ and $\vert \phi_{-} \rangle $ via single-quantum transitions, with relaxation rate scaling as $ \cos^2 \theta $; and via double quantum transitions, with relaxation rate scaling as $ \cos^2 \theta \sin^2 \theta $.

The above scaling of the relaxation of $\rho_{ST}$ suggests a model for the measured singlet lifetime as a function of RF spin-locking power:
\begin{align}
\frac{1}{T_{S,measured}} =& \frac{1}{T_x} \cos^2 \theta + \frac{1}{T_S}\\
=& \frac{1}{T_x} \frac{1}{1+(2\nu_n / \Delta \nu)^2} + \frac{1}{T_S}   .
\end{align}
where $T_x$ is the lifetime at low RF power and $T_S$ is the maximum singlet lifetime, typically achieved at high RF power. Significantly, this model predicts that for typical maximum singlet lifetimes, the measured singlet lifetime reaches $95\%$ of its maximum value when the nutation rate $\nu_n$ is approximately 5 $\Delta \nu$.

Relaxation of the long-lived coherence $\rho_{LLC}$ can be modeled using a similar analysis. In terms of the spin-locked eigenstates, we have:
\begin{align}
\rho_{LLC} =& \vert \uparrow \downarrow \rangle \langle \uparrow \downarrow \vert - \vert \downarrow \uparrow \rangle \langle \downarrow \uparrow \vert\nonumber \\ =& \cos \theta (\vert \phi _{+}\rangle \langle \phi _{+}\vert-\vert \phi _{-}\rangle \langle \phi _{-}\vert)\nonumber\\+&\frac{ \sin \theta}{\sqrt{2}} (\vert \phi _{S}\rangle \langle \phi _{+}\vert+\vert \phi _{+}\rangle \langle \phi _{S}\vert+\vert \phi _{S}\rangle \langle \phi _{-}\vert+\vert \phi _{-}\rangle \langle \phi _{S}\vert)   .
\end{align}

At low RF spin-locking powers, the long-lived coherence is mainly composed of $\vert \phi _{+}\rangle \langle \phi _{+}\vert-\vert \phi _{-}\rangle \langle \phi _{-}\vert$, and these two eigenstates interact with one another via a zero-quantum transition. However, as the RF power is increased, these states begin to mix with $\vert T_{+} \rangle$ and $\vert T_{-} \rangle$, which opens up double-quantum transitions with relaxation rates scaling as $\sin^4 \theta$. A double-quantum transition with $\vert \phi_0 \rangle $ also becomes available, with relaxation rate scaling as $\sin^2 \theta $. The latter relaxation rate increases more quickly with RF power and dominates at small $\theta $. 

This above scaling suggests a simple model for the $\rho_{LLC}$ relaxation rate:
\begin{align}
\frac{1}{T_{LLC,measured}} =& \frac{1}{T_y} \sin^2 \theta + \frac{1}{T_{LLC}}\\
=& \frac{1}{T_y} \frac{(2\nu_n / \Delta \nu)^2}{1+(2\nu_n / \Delta \nu)^2} + \frac{1}{T_{LLC}}   .
\end{align}
where $1/T_{LLC}$ is the relaxation rate at zero RF power and $1/T_y$ is the additional relaxation rate due to the applied RF power.

We find that our model for the measured singlet lifetime agrees well with the detailed treatment of Pileio and Levitt (see eq. 43 in \cite{Pileio1}), which contains terms up to eighth power in $\cos \theta$. Our model includes only lowest-order terms, but satisfactorily describes the measured relationship between singlet lifetime and RF power, as described below. The two models deviate most at low RF powers ($\nu_n < 0.5 \Delta \nu$), where higher-order terms in $\cos \theta$ make larger contributions. See supplement S2 for a comparison.

\section{Experimental Results}

We performed NMR studies at 4.7 T of proton pair singlet states in a number of small organic molecules using a wide range of RF spin-locking powers. We chose citric acid and p-hydroxybenzoic acid, as Pileio {\em et al.} had previously studied these using high RF power \cite{Levitt1}. Aditionally, we studied aspartic acid, trans-1,4-cyclohexanediol, and glycerol formal as examples of molecules with a range of structures. 

\begin{figure}[h!]
\vspace*{.05in}
\centering
\includegraphics[scale=1]{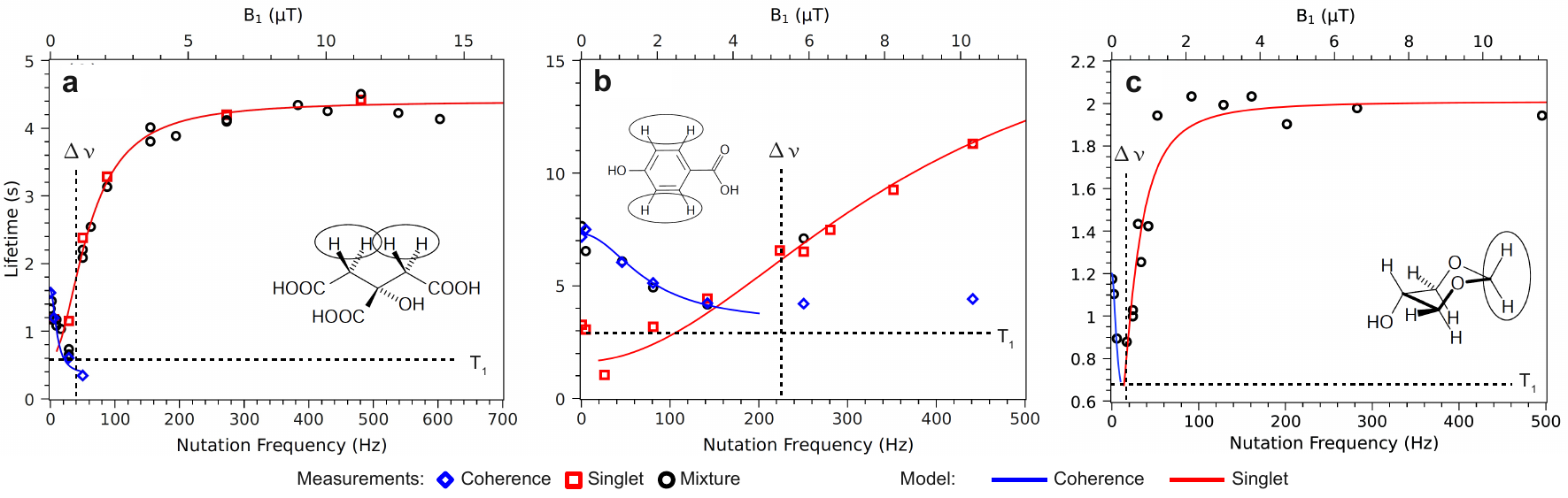}  
\caption{Measurements of the lifetimes of the singlet state, long-lived coherence (LLC), and a mixture of the two as a function of the effective RF spin-locking field $B_1$ for proton pairs in small organic molecules. Also shown are fits to models for the singlet and LLC lifetimes, as described in the main text. (A) Citric acid, $T_x = 500$ ms and $T_y = 600$ ms; (B) p-hydroxybenzoic acid $T_x = 2$ s and $T_y = 7$ s; (C) glycerol formal, $T_x = 250$ ms and $T_y = 1$ s; RF power is quantified by the induced nutation frequency about the $B_1$ field. Molecular structures are shown, protons of the singlet(s) are circled, and values for chemical shifts ($\Delta \nu$) and spin-lattice relaxation times ($T_1$) are indicated.}
\end{figure}

Our experimental protocol (Fig. 2) initialized proton pairs into one of three different states: the singlet-enhanced superposition state $\rho_{ST}$, which contained predominantly singlet population after rapid initial decay of the triplet component; a long-lived coherence between singlet and triplet, $\rho_{LLC}$; or a mixture of the two.
In all molecules we measured the singlet ($\rho_{ST}$) lifetime to increase with the applied RF power, reaching a plateau at the maximum singlet lifetime, $T_S$, when $\nu_n > 5\Delta \nu$. In contrast, we found that the lifetime of the long-lived coherence decreases from its maximum value of $T_{LLC}$ with the application of RF power. Both of these cases are well-modeled by equations 18 and 21 above. When we created a mixture of $\rho_{ST}$ and $\rho_{LLC}$, the measured lifetime was that of the state with the longest lifetime at a given RF power, although the measured amplitude was lower as the contribution from the faster-relaxing state was quickly lost. We individually fit the two regimes of the mixed-state lifetime-vs.-RF-power measurements with the corresponding models for $\rho_{ST}$ and $\rho_{LLC}$, which provided a good characterization of the system's behavior, as shown in Fig. 3A-C. Note that for each molecule studied, we found that the shortest mixed-state lifetime occurs near $\nu_n \approx 0.5 \Delta \nu$. Results for maximum singlet and LLC lifetimes ($T_S$ and $T_{LLC}$) are summarized in Table 1.

\begin{table}[h!]
\centering
\caption{Measured values of spin-lattice, singlet, and long-lived coherence (LLC) relaxation times.}
\begin{tabular*}{\hsize}{@{\extracolsep{\fill}}lccccc}
Molecule & $T_1 (s)$ & $T_S (s)$ & Enhancement ($T_S/T_1$) & $T_{LLC} (s)$ & Enhancement ($T_{LLC}/T_1$)\cr
\hline
citric acid               & $0.58 \pm 0.03$ & $4.5 \pm 0.3 $   & $7.8 \pm 0.7 $ & $1.5 \pm 0.1 $   & $2.6 \pm 0.2$\cr
p-hydroxybenzoic acid D$_{2}$O & $2.9 \pm 0.1$   & $16 \pm 2 $ & $6.2 \pm 0.8   $ & $7.3 \pm 0.7 $   & $2.5 \pm 0.3$\cr
p-hydroxybenzoic acid H$_{2}$O & $2.3 \pm 0.1$   & $5.8 \pm 0.2 $ & $2.5 \pm 0.1  $ & $3.9 \pm 0.1 $   & $1.7 \pm 0.1$\cr
aspartic acid             & $0.83 \pm 0.03$ & $7.48 \pm 0.3 $  & $9.0 \pm 0.5 $ & $2.3 \pm 0.4 $   & $2.8 \pm 0.5$\cr
trans-1,4-cyclohexanediol & $1.34 \pm 0.02 $& $3.9 \pm 0.9 $& $2.9 \pm 0.7   $       & $2.90 \pm 0.01 $ & $ 2.16 \pm 0.03$\cr
glycerol formal           & $0.68 \pm 0.01$ & $1.91 \pm 0.03 $ & $2.81 \pm 0.06 $ & $1.37 \pm 0.02 $ & $2.01 \pm 0.04$\cr
\end{tabular*}
\end{table}

We also investigated possible mechanisms for proton-pair singlet relaxation. First, we compared our measurements with those at other magnetic fields to probe the importance of chemical-shift anisotropy (CSA) relaxation (see Table 1). Our result of $T_S = 4.5$ s and 7.8-fold lifetime enhancement over $T_1$ for citric acid at 4.7 T are consistent with a previous measurement of 4.81 s and 7.6 $T_1$ at 9.4 Tesla \cite{Levitt1}. Due to hardware limitations, there was insufficient RF power to reach the maximum singlet lifetime for p-hydroxybenzoic acid. Nevertheless, a fit to data at finite RF power gave $T_S$ = 16 s and 5.5 $T_1$, which is consistent with previous measurements \cite{Levitt1}. Since these singlet-state lifetimes do not significantly depend on magnetic field strength, we conclude that CSA is not a primary mechanism of singlet relaxation. To distinguish between the spin-rotation and magnetic quadrupole relaxation mechanisms, we performed singlet lifetime measurements at several sample temperatures. We found that the singlet lifetime increases linearly with temperature, which identifies the magnetic quadrupole relaxation mechanism as dominant in such proton-pair singlet state molecules (see supplement S3).

Note that most previous proton-pair singlet measurements were conducted using deuterated solvents, which should result in weaker singlet-solvent interactions and larger enhancements of singlet state lifetime. To test whether such lifetime enhancement changed in a normally protonated solvent, we studied p-hydroxybenzoic acid in both D$_{2}$O and H$_{2}$O. We found that the enhancement of both the singlet and LLC lifetimes were significantly lower in H$_{2}$O (see Table 1). The enhancement is likely higher in D$_{2}$O due to the substitution of deuterium for the phenolic proton as well as reduced dipolar interactions with nearby solvent protons. 

\section{Discussion}
The above experimental results and associated modeling establish an operational spin-locking condition $\nu_n^{opt} \approx 5 \Delta \nu$ to realize maximum singlet lifetime with minimal RF power. In the context of this operational condition, we can reassess the past work by Levitt and colleagues using high-power RF spin-locking fields \cite{Levitt1,Levitt6,Levitt9}. As shown in Table 2, most of the previous experiments employed $\nu_n >> 5 \Delta \nu$: i.e., they used much higher RF power than was needed to achieve a long singlet-state lifetime. For example, for citric acid Pileio {\em et al.} \cite{Levitt1} used $\nu_n = 3.5$ kHz, whereas $\Delta \nu = 72$ Hz at 9.4 T, which is an order-of-magnitude higher spin-locking field than necessary.

Furthermore, we note that similarly low RF powers will be required for practical {\em in vivo} singlet-state creation in a wide variety of molecules using clinical MRI scanners, where the static magnetic field is commonly between 1.2 and 7 T; see example values for $\nu_n^{opt}$ at 1.5 T given in Table 2. For example, at 4.7 T glycerol formal's protons have a frequency difference $\Delta \nu \approx 16$ Hz; hence $\nu_n^{opt} \approx 80$ Hz is sufficient to achieve significant singlet-state lifetime enhancement. For common biomolecules such as citric acid and aspartic acid, $\nu_n^{opt} <100$ Hz at 1.5 T, which is well within the spin-locking regime commonly used in clinical MRI \cite{Reddy1,Bachert1}. The spin-locking times and strengths used in \cite{Reddy1,Bachert1} imply that a 60 Hz spin-lock could be safely applied for 3.5 s, and a 20 Hz spin-lock for 30 s. These timescales are of the same order as the singlet lifetimes we measured for typical small molecules, and thus should be sufficient to conduct a variety of {\em in vivo} measurements using singlets. Alternatively, the long-lived coherence can be utilized without the need for any spin-locking if only moderate lifetime enhancements are required.

In summary, our measurements and theoretical description show that for many molecules long-lived nuclear-spin singlet states and singlet/triplet coherences can be created using RF spin-locking powers that are more than two orders of magnitude lower than in previous studies; and that the effectiveness of the spin-locking can be accurately predicted from spectral parameters. These insights will be useful in the development of new applications for singlet states {\em in vivo}, where the RF specific-absorption rate (SAR) must be minimized.

\begin{table}[h!]
\centering
\caption{Comparison of the chemical shift; optimal spin nutation frequency for RF spin-locking, $\nu_n^{opt} \approx 5\Delta \nu$; and values of $\nu_n$ used in previous experiments. Also listed are values for $\nu_n^{opt}$ for a clinical MRI scanner.} 
\begin{tabular*}{\hsize}{lcccc}
Molecule & $\Delta \nu $ ($B_0$ field) & $\nu_n^{opt}$ & $\nu_n$ used in previous experiments & $\nu_n^{opt}$ at 1.5 T \cr
\hline
citric acid \cite{Levitt1} & 72 Hz (9.4 T) & 360 Hz & 3500 Hz & 57 Hz \cr
p-hydroxybenzoic acid \cite{Levitt1} & 445 Hz (11.75 T) & 2224 Hz & 3500 Hz & 284 Hz \cr
aspartic acid \cite{Bodenhausen5} & 100 Hz (11.75 T) & 500 Hz  & 2500 Hz & 64 Hz 
\end{tabular*}                   
\end{table}

\section{Materials and Methods}
Solutions of citric acid, aspartic acid, p-hydroxybenzoic acid, and 1,4-cyclohexanediol were made in D$_2$O, with the addition of sodium hydroxide where necessary for dissolution. Glycerol formal was analyzed neat. Concentrations and conditions can be found in Table 3. All reagents were purchased from Sigma-Aldrich . All samples were prepared in 10 mm diameter NMR sample tubes and bubbled with nitrogen gas for three minutes. Spectra were acquired on a 200 MHz Bruker spectrometer without spinning.

Experiments shown in Fig. 2 were run on each compound using varying lengths for $\tau_4$. Pulse sequence parameters can be found in Table 4. To remove any remaining triplet polarization, phase cycling was used in which the experiment was repeated with both the first and last 90$^{\circ}$ pulses along -x rather than x. Between 8 and 32 averages were used to provide sufficient signal-to-noise. The intensity of each peak was then measured and plotted against $\tau_4$. The resulting data was fit with a single exponential time decay. Further details of the experimental pulse sequences are discussed in Supplement S4. Multiple datasets were collected using different RF power levels for spin-locking. The RF power was characterized by measuring the nutation rate induced by the RF $B_1$ field, which was calibrated using a sequence of single-pulse experiments performed with increasing pulse length.
$T_1$ relaxation rates were measured through conventional inversion-recovery experiments.

For variable temperature experiments, the temperature was controlled by supplying hot air to the probehead. Blown air was heated with a Hotwatt cartridge heater controlled by an Omron temperature control box, and the temperature of the sample was monitored with an RTD in the probehead.

\begin{table}[h!]
\centering
\caption{Sample preparations for the study of long-lived states.}
\begin{tabular*}{\hsize}{lccl}
Molecule & Concentration & NaOH Concentration & Solvent\cr
\hline
citric acid & 0.26 M & 0 & D$_{2}$O\cr
p-hydroxybenzoic acid & 0.29 M & 0.50 M & D$_{2}$O, H$_{2}$O\cr
aspartic acid & 0.020 M & 1.0 M & D$_{2}$O\\
1,4-cyclohexanediol (cis/trans mixture) & 0.41 M & 0 & D$_{2}$O\cr
glycerol formal & neat & 0 & neat \cr
\end{tabular*}
\end{table}

\begin{table}[h!]
\centering
\caption{Delays, in ms, for pulse sequences used in the experiments: $\tau_1-\tau_2- \tau_3-\tau_4-\tau_5$}
\begin{tabular*}{\hsize}{@{\extracolsep{\fill}}lccl}
Molecule & Singlet & Coherence & Mixture\cr
\hline
citric acid & 12.5-3.7-6.0-$\tau_4$-6.0 & 12.0-17.0-$\tau_4$-15.5 & 12.0-12.0-$\tau_4$-12.0\cr
p-hydroxybenzoic acid & 29.3-31.5-1.14-$\tau_4$-1.14 & 2.0-3.75-$\tau_4$-1.25 & 30.0-1.25-$\tau_4$-1.25\cr
aspartic acid & 11.8-8.0-3.5-$\tau_4$-4.0 & - & 7.4-7.0-$\tau_4$-3.7\cr
1,4-cyclohexanediol (cis/trans mixture) & - & - & 4.2-3.0-$\tau_4$-2.0\cr
glycerol formal & - & - & 31.0-15.0-$\tau_4$-15.0 \cr
\end{tabular*}
\end{table}

\newpage
\bibliographystyle{model1a-num-names}
\bibliography{singlet_power_bib}
%

\end{document}